\documentclass[twocolumn,preprintnumbers,superscriptaddress,amsmath,amssymb,prl]{revtex4-2}
\usepackage{amsmath,amssymb,graphicx}
\usepackage{color}
\usepackage{siunitx}
\usepackage{physics}
\usepackage{circuitikz}
\usepackage{float}
\usepackage{bm}

\begin{document}

\title{Switchable Non-Hermitian Skin Effect in Bogoliubov Modes}
\newcommand{\thedate}{Sep 2024}

\author{Hsuan Lo}
\affiliation{School of Physical and Mathematical Sciences, Nanyang Technological University,
Singapore 637371, Singapore}

\author{You Wang}
\affiliation{School of Physical and Mathematical Sciences, Nanyang Technological University,
Singapore 637371, Singapore}

\author{Rimi Banerjee}
\affiliation{School of Physical and Mathematical Sciences, Nanyang Technological University,
Singapore 637371, Singapore}

\author{Baile Zhang}
\email{blzhang@ntu.edu.sg}
\affiliation{School of Physical and Mathematical Sciences, Nanyang Technological University, Singapore 637371, Singapore}
\affiliation{Centre for Disruptive Photonic Technologies, Nanyang Technological University, Singapore, 637371, Singapore}

\author{Y. D. Chong}
\email{yidong@ntu.edu.sg}
\affiliation{School of Physical and Mathematical Sciences, Nanyang Technological University, Singapore 637371, Singapore}
\affiliation{Centre for Disruptive Photonic Technologies, Nanyang Technological University, Singapore, 637371, Singapore}

\begin{abstract}
  Interacting or nonlinear lattices can host emergent particle-like modes, such as Bogoliubov quasiparticles, whose band topology and other properties are potentially highly tunable.  Despite originating in the study of superconducting materials, Bogoliubov quasiparticles can also occur in synthetic metamaterials.  Here, we implement a nonlinear driven-dissipative circuit whose fluctuations are Bogoliubov modes possessing nontrivial non-Hermitian band topology.  We show experimentally that the system exhibits a switchable non-Hermitian skin effect (NHSE), which abruptly appears when the on-site driving voltage amplitude exceeds a threshold.  In contrast to earlier realizations of the NHSE and related phenomena in circuit models, the switchable NHSE in our system occurs in Bogoliubov modes, which are strongly affected by how the system is driven.  Moreover, unlike other experimental platforms hosting non-Hermitian Bogoliubov modes, our system does not rely on $p$-wave-like pairing interactions, only local nonlinearities.
\end{abstract}

\maketitle

%%% intro

Interacting phases of matter can host emergent quasiparticles that can potentially possess nontrivial band topology.  Such quasiparticles, if they can be reliably produced, may have important implications; for instance, Majorana quasiparticles in topological superconductors \cite{alicea2012new, beenakker2013search, sato2016majorana, sato2017topological} could enable robust quantum computing \cite{aasen2016milestones, sarma2015majorana}.  The realization of topological superconductors is a major ongoing challenge \cite{fu2008superconducting, leijnse2012introduction, sato2017topological, zhang2018observation, frolov2020topological, trang2020conversion}, but some of the underlying ideas can be explored using nonlinear synthetic metamaterials \cite{ozawa2019rmp}.  The Bogoliubov-de Gennes (BdG) transformation \cite{bogoliubov1947theory,zhu2016bogoliubov}, the mechanism giving rise to Bogoliubov quasiparticles in topological superconductors, is also applicable to interacting bosonic systems; however, the BdG Hamiltonian for bosonic systems is not guaranteed to be Hermitian \cite{lieu2018topological, McDonald2018, banerjee2020coupling, xu2021interaction, Wang2022nhse}, so the Bogoliubov quasiparticles can have qualitatively different properties, including a broader set of topological phases  \cite{bender2007making, nakamura2008condition, kawaguchi2012spinor, kawabata2019symmetry, ashida2020non, bergholtz2021exceptional, Wang2023tutorial}.  For example, in exciton-polariton metamaterials \cite{deng2010rmp, carusotto2013rmp, banerjee2020coupling}, Bogoliubov quasiparticles can exhibit the non-Hermitian skin effect (NHSE) \cite{xu2021interaction}, which is tied to a form of nontrivial non-Hermitian (NH) band topology called point gap winding \cite{hatano1996localization, yao2018edge, bergholtz2021exceptional, zhang2022review}.  This behavior has recently been demonstrated in experiments on optomechanics \cite{slim2024optomechanical} and superconducting quantum circuits \cite{busnaina2024quantum}.

Here, we experimentally realize NH Bogoliubov quasiparticles using a nonlinear electric circuit, and uncover an actively-switchable form of the NHSE different from the fixed skin effects of earlier models.  When the amplitude of a driving field increases above a certain threshold, the Bogoliubov modes abruptly switch from trivial to nontrivial point gap topology, resulting in a sudden onset of localization in a sideband.  Our experiment uses an RLC transmission line \cite{albert2015topological,ningyuan2015time,lee2018topolectrical,imhof2018topolectrical,ezawa2019electric,hofmann2019chiral,helbig2020generalized,hofmann2020reciprocal,liu2021non} containing varactors (nonlinear capacitors) \cite{hadad2018self,wang2019topologically}, which maps onto a one-dimensional nonlinear lattice hosting NH Bogoliubov quasiparticles.  While linear circuits have been extensively used to study the NHSE and other aspects of band topology \cite{luo2018nodal, ezawa2019electric, hofmann2019chiral, ezawa2019electric, helbig2020generalized, liu2020gain, zhang2020non}, the switchable NHSE explored in this work relies on both non-Hermiticity and nonlinearity, with the Bogoliubov quasiparticles manifesting as sideband modes away from the fundamental driving frequency.  We note that there have been previous experiments on nonlinear circuit lattices focusing on quasiparticles produced via higher-harmonic generation, which is distinct from the Bogoliubov mechanism \cite{buryak1995solitons, shadrivov2006second, shadrivov2008tunable, Wang2019nonlin, smirnova2019third}.

The Bogoliubov quasiparticles in our system are fluctuations on an externally-driven fundamental mode \cite{banerjee2020coupling, xu2021interaction}, governed by a NH BdG Hamiltonian.  The BdG Hamiltonian possess a pseudo-Hermitian symmetry \cite{mostafazadeh2002pseudo}, whose spontaneous breaking enables switchable point gap topology.  Under weak driving, the pseudo-Hermiticity is unbroken and all eigenvalues are real \cite{bender2007making, ashida2020non}, so there is no point gap; after crossing a driving threshold, the pseudo-Hermiticity spontaneously breaks, inducing a point gap and skin modes \cite{bergholtz2021exceptional}.  This is unlike past realizations of linear and nonlinear NHSEs, since the underlying circuit lacks the asymmetric inter-site hoppings that normally play a key role in manifesting the NHSE \cite{hatano1996localization, yao2018edge, hofmann2020reciprocal, helbig2020generalized, liu2021non}.  In particular, our model differs from the bosonic Kitaev chain \cite{mcdonald2018phase} implemented in the aforementioned optomechanics and quantum circuit experiments \cite{slim2024optomechanical, busnaina2024quantum}, a nonlinear model that relies on a nonlocal $p$-wave pairing interaction to generate effective asymmetric hoppings.  By contrast, our scheme only requires a local Kerr-like nonlinearity; the NH Bogoliubov quasiparticles occur as sideband modes, and the fundamental drive is used to impart the left-right asymmetry to the BdG Hamiltonian that produces the switchable point gap and NHSE.

%%% result

\begin{figure}
    \centering
    \includegraphics[width=\linewidth]{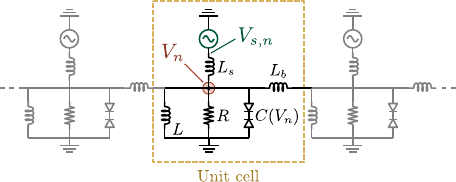}
    \caption{Design of a driven nonlinear transmission line circuit with a similar BdG Hamiltonian.  The dashed box indicates unit cell $n$.  Each unit cell contains a nonlinear capacitance $C(V_n)$ implemented by back-to-back varactors.}
    \label{fig:1_schematic}
\end{figure}

\textit{Toy model}---Before presenting the experiment, we illustrate its key features by reviewing a toy model \cite{carusotto2013rmp,kevrekidis2009discrete}.  Consider a driven-dissipative Gross-Pitaevskii equation for two sites $\sigma \in \{0,1\}$,
\begin{equation}
    i\pdv{\psi_\sigma}{t} = (\varepsilon - i\gamma)\psi_\sigma + J \psi_{\sigma'} + \frac{g}{2}|\psi_\sigma|^2\psi_\sigma + F_p e^{-i\omega_p t + ik_p\sigma},
    \label{GP_simple}
\end{equation}
where $\psi_\sigma$ is the complex amplitude, $\sigma' \neq \sigma$ is the opposite site, $\varepsilon$ and $\gamma$ are the on-site frequency and loss, $J$ is the inter-site hopping, and $g$ is a Kerr coefficient.  A coherent drive is applied with amplitude $F_p$, frequency $\omega_p$, and phase difference $k_p$ between the sites. 

In the weakly nonlinear regime $g|\psi_\sigma|^2 \ll \varepsilon$, we can apply the BdG transformation \cite{banerjee2020coupling, xu2021interaction} by taking the ansatz
\begin{equation}
  \psi_\sigma = \left(\Psi\, e^{ik_p \sigma} + u_\sigma e^{-i\omega_f t}
  + v_\sigma^* e^{i\omega_f^*t} \right) e^{-i\omega_pt}.
  \label{toy_ansatz}
\end{equation}
Plugging this into \eqref{GP_simple} yields the eigenproblem
\begin{equation}
    \label{eq:eigenproblem}
    \mathbf{H}
    \begin{pmatrix}
        \mathbf{u} \\
        \mathbf{v} \\
    \end{pmatrix}
    = \tilde{\omega}_f
    \begin{pmatrix}
        \mathbf{u} \\
        \mathbf{v} \\
    \end{pmatrix},
\end{equation}
where $\tilde{\omega}_f = \omega_f + i\gamma$, $\mathbf{u} = (u_0, u_1)^T$, $\mathbf{v} = (v_0, v_1)^T$, and
\begin{align}
    \label{eq:minimal H}
    \mathbf{H} &= 
    \begin{pmatrix}
        \mathbf{H}_u & \mathbf{H}_v \\
        -\mathbf{H}_v^* & -\mathbf{H}_u
    \end{pmatrix}, \\
    \mathbf{H}_u
    &= \left(\varepsilon - \omega_p + g|\Psi|^2\right) \mathbf{I}
    + J\bm{\sigma}_1, \\
    \mathbf{H}_v &= \frac{g|\Psi|^2}{4}  \left[ (1 + e^{2ik_p}) \mathbf{I}
      + (1 - e^{2ik_p}) \bm{\sigma}_3 \right], \label{Hv}
\end{align}
where $\mathbf{I}$ is the $2\times2$ identity matrix, $\bm{\sigma}_i$ is the $i$-th Pauli matrix, and $\omega_f$ is the BdG eigenfrequency. Since $\mathbf{H}_v$ is diagonal, the BdG Hamiltonian $\mathbf{H}$ is manifestly NH. For $k_p \ne 0$, $\mathbf{H}_v$ introduces asymmetric couplings between the $u$ and $v$ sectors, which act like nonreciprocal lattice hoppings \cite{xu2021interaction}.

This BdG Hamiltonian obeys a pair of non-Hermitian symmetries \cite{mostafazadeh2002pseudo, nakamura2008condition, schnyder2008classification, lieu2018topological, kawabata2019symmetry, Wang2023tutorial}: pseudo-Hermiticity, $\mathbf{\Gamma}^3 \mathbf{H} \mathbf{\Gamma}^3 = \mathbf{H}^\dagger$, and non-Hermitian particle-hole (NHPH) symmetry, $\mathbf{\Gamma}^1 \mathbf{H}^* \mathbf{\Gamma}^1 = -\mathbf{H}$, where $\mathbf{\Gamma}^i = \bm{\sigma}_i \otimes \mathbf{I}$.  Consequently, the BdG eigenfrequencies form a quadruplet $\{\tilde{\omega}_f, -\tilde{\omega}_f^*, -\tilde{\omega}_f, \tilde{\omega}_f^*\}$ if both symmetries are spontaneously broken, or two pairs $\{\tilde{\omega}_f, -\tilde{\omega}_f\}$ and $\{\tilde{\omega}_f^\prime, -\tilde{\omega}_f^\prime\}$ that are purely real (if pseudo-Hermiticity is unbroken) or imaginary (if NHPH symmetry is unbroken).  Another consequence is that an eigenstate which spontaneously breaks pseudo-Hermicity must satisfy $\lVert \mathbf{u}\rVert^2 = \lVert \mathbf{v}\rVert^2$, where $\lVert \mathbf{w}\rVert^2 \equiv \sum_n |w_n|^2$ (see Supplemental Materials \cite{SM}).  This property, which we call ``non-Hermitian particle/hole pinning'', will be useful in the experiment.

\textit{Nonlinear circuit}---A BdG Hamiltonian of the above form can be realized via the transmission line circuit shown schematically in Fig.~\ref{fig:1_schematic}.  This is an lattice of inductively-coupled RLC resonators, each containing a pair of back-to-back varactors that implement a Kerr-like nonlinearity \cite{hadad2018self, wang2019topologically}.  For small voltages ($V \lesssim 1\,\textrm{V}$), the nonlinear capacitance has the form $C(V) \approx C_0 + C_1V^2$. The other transmission line parameters are the on-site resistance $R$, on-site inductance $L$, and binding inductance $L_b$.  Each site $n$ is coupled by an inductor $L_s$ to an AC voltage source $V_{s,n} = \frac{1}{2}V_{pp}\sin{(\omega_p t + k_p n)}$, where $V_{pp}$ and $\omega_p$ are tunable and $k_p = 2\pi/3$.

Away from the boundaries, the voltage at resonator $n$, denoted by $V_n$, obeys the equation of motion
\begin{multline}
    C_0\Ddot{V_n} + \frac{\Dot{V_n}}{R} + \left(\frac{1}{L_s} + \frac{2}{L_b} + \frac{1}{L} \right)V_n + C_1\Ddot{V_n}V_n^2 \\ 
    + 2C_1\Dot{V_n}^2V_n
    + \frac{1}{L_b} \left(V_{n+1} + V_{n-1}\right)
    + \frac{V_{s,n}}{L_s} = 0.
    \label{eq:EOM}
\end{multline}
We apply the slowly-varying envelope approximation using the ansatz
\begin{equation}
    \label{eq:SVEA}
  V_n = \frac{1}{2}\left(\Phi_n(t) \, e^{-i\omega_p t} + \,\mathrm{c.c.} \right),
\end{equation}
followed by a BdG ansatz for the envelope function:
\begin{equation}
    \label{eq: BdG transformation}
    \Phi_n(t) = \Psi e^{ik_pn} + u_n \, e^{-i\omega_f t} + v_n^* \, e^{i\omega_f^* t}.
\end{equation}

It can then be shown \cite{SM} that $u_n$ and $v_n$ obey a BdG eigenproblem.  The BdG Hamiltonian has the same form as \eqref{eq:minimal H}, with $\mathbf{H}_u$ and $\mathbf{H}_v$ expressed in terms of various circuit parameters. For convenience, we will normalize the BdG eigenfrequencies to make them dimensionless, and denote them by $\Omega_f$; the corresponding physical frequency is $\Omega_f\omega_p - i \gamma$, where $\gamma$ is a reference decay constant. For a finite lattice under Dirichlet-like or ``open'' boundary conditions (OBC), the BdG Hamiltonian obeys pseudo-Hermiticity and NHPH symmetry, with the same consequences as in the toy model (with $\Omega_f$ replacing $\tilde{\omega}_f$ in the eigenvalue relations).  For Floquet-Bloch periodic boundary conditions (PBC) with quasimomentum $k = k_p \neq 0$, pseudo-Hermiticity remains but the NHPH symmetry is explicitly broken.

\begin{figure}
    \centering
    \includegraphics[width=\linewidth]{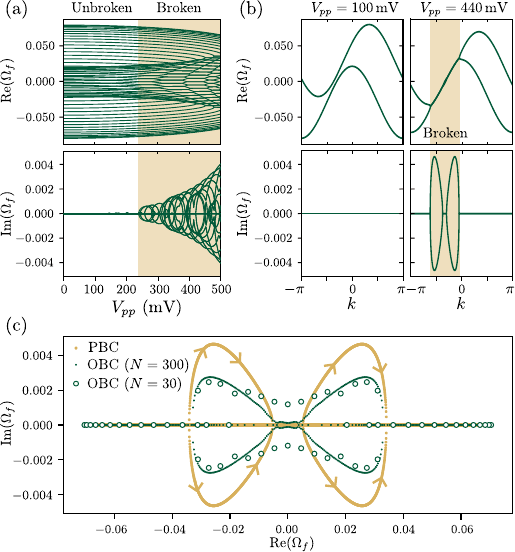}
    \caption{(a) Bogoliubov eigenfrequency spectrum calculated for the circuit with $N = 30$ unit cells, with OBC, under different driving voltage amplitudes. Pseudo-Hermiticity spontaneously breaks when $V_{pp} \gtrsim \SI{240}{mV}$.  The yellow region indicates the regime of spontaneously broken pseudo-Hermiticity.  (b) Bogoliubov band spectrum under PBC for $V_{pp} = \SI{100}{mV}$ (left panels) and $V_{pp} = \SI{440}{mV}$ (right panels).  (c)  Complex spectra calculated at $V_{pp} = \SI{440}{mV}$.  For the PBC case, each band forms a pair of loops connected by straight segments (yellow curves, with arrows indicating the direction of increasing $k$ as it sweeps from $-\pi$ to $\pi$).  The eigenfrequencies under OBC (dots: $N = 300$; circles: $N = 30$) deviate markedly from the loop trajectories of the PBC case.}
    \label{fig:2_theory}
\end{figure}

\begin{figure*}
    \centering
    \includegraphics[width=\linewidth]{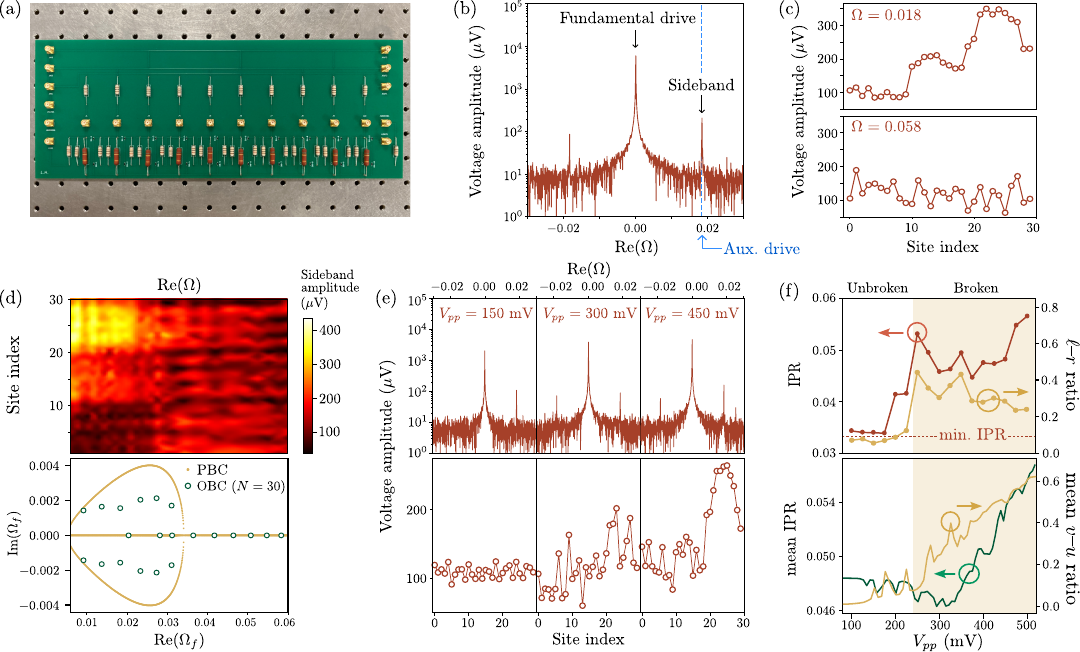}
    \caption{(a) Circuit implementation of BdG modes with actively switchable NHSE.  The sample consists three identical printed circuits boards (PCBs), one of which is shown here, containing 10 sites each.  (b) A measured voltage spectrum, at site $n = 30$ for $V_{pp} = \SI{500}{mV}$.  (c) Distribution of sideband voltage amplitudes for $\Omega=0.018$ (top panel) and $\Omega=0.058$ (lower panel).  The former exhibits strong localization, indicative of the NHSE. (d) Upper panel: sideband amplitude distribution at different auxiliary drive frequencies. Lower panel: calculated Bogoliubov spectra under OBC and PBC.  (e) Upper panels: voltage spectra at $n = 30$ for different fundamental drive amplitudes. Lower panels: the corresponding sideband voltage amplitude distributions.  (f) Upper panel: experimental results for the inverse participation ratio (IPR, red line) and left-right sideband ratio $\lVert\mathbf{V}^\ell\rVert/\lVert\mathbf{V}^r\rVert$ (yellow line) versus driving amplitude. Lower panel: theoretical results for the IPR (green line) and particle/hole ratio $\lVert\mathbf{v}\rVert/\lVert\mathbf{u}\rVert$ (yellow line), averaged over all BdG modes, versus driving amplitude.  The yellow-shaded region shows the regime of spontaneously broken pseudo-Hermiticity from Fig.~\ref{fig:2_theory}(a).  In all subplots, the fundamental frequency is $\omega_p/2\pi = \SI{3.81}{MHz}$ and the auxiliary drive amplitude is $V_{pp}^\prime = \SI{50}{mV}$.  In (b) and (e)--(f), the auxiliary drive frequency is fixed at $\omega^\prime/2\pi = \SI{3.88}{MHz}$, corresponding to $\Omega = 0.018$. }
    \label{fig:3_exp}
\end{figure*}

These properties set the stage for an actively switchable NHSE.  In Fig.~\ref{fig:2_theory}(a), we plot the spectrum of Bogoliubov eigenfrequencies $\Omega_f$ for the finite ($N = 30$) lattice under OBC, under different driving voltage amplitudes. For $V_{pp} \lesssim \SI{240}{mV}$, the vast majority of eigenmodes have unbroken pseudo-Hermiticity and hence real $\Omega_f$ \cite{bender2007making}. As $V_{pp}$ is increased past this threshold, the pairwise pseudo-Hermiticity breaks spontaneously and quadruplets with non-real $\Omega_f$ start appearing.  (There are also isolated instances of pseudo-Hermiticity breaking over small intervals of $V_{pp}$ before the $\SI{240}{mV}$ threshold; the number of such modes is $\ll N$, so this appears to be a finite-size effect.) Note that we have defined $\Omega_f$ so that the physical frequencies have an imaginary shift relative to it; in the experimental regime, the physical frequencies have negative imaginary parts (i.e., the modes are dissipative).  The NHPH symmetry is spontaneously broken throughout.

In Fig.~\ref{fig:2_theory}(b), we plot the complex band spectrum for the Bogoliubov quasiparticles under PBC.  There are two bands due to the enlargement of the Hamiltonian under the BdG transformation.  In the large $V_{pp}$ regime, we see that the breaking of pseudo-Hermiticity does not occur everywhere in the Brillouin zone, but only a certain range of $k$.  Thus, in the complex $\Omega_f$ plane, the spectrum forms the curves plotted in yellow in Fig.~\ref{fig:2_theory}(c), consisting of two pairs of loops combined with straight segments lying on the real axis.  Each band forms a point gapped spectrum, with winding number $\pm 1$ around any reference energy within any of the loops \cite{zhang2020correspondence}.  When the finite system's eigenfrequencies (under OBC) are plotted in the same graph, they are found to deviate from the PBC eigenfrequencies, as shown by the dots (for $N = 300$) and circles (for $N = 30$) in Fig.~\ref{fig:2_theory}(c). These features---point-gap winding under PBC and lack of correspondence with the OBC spectrum---are characteristic of the NHSE \cite{hatano1996localization, yao2018edge, bergholtz2021exceptional, zhang2022review}.  Specifically, the OBC eigenfrequencies form segmented curves in the complex plane, with $\mathrm{Im}(\Omega_f) \ne 0$, that are encircled by the PBC eigenfrequencies; these are skin modes, localized to one side of the lattice determined by the winding of the PBC eigenfrequency loop enclosing them.  The eigenmodes with negative (positive) $\Im(\Omega_f)$ have anticlockwise (clockwise) winding and are localized to the left (right) \cite{zhang2020correspondence}.  The presence of paired complex eigenfrequencies, which share the same real part but lie on loops with opposite windings, is a feature observed in previous models and dubbed the ``double-sided skin effect'' \cite{xu2021interaction}; however, the lossier set of skin modes is not observable in our present experiment \cite{SM}.

\textit{Results}---We implement the transmission line circuit on printed circuit boards (PCBs) such as those shown in Fig.~\ref{fig:3_exp}(a).  The circuit component parameters are $R=\SI{150}{k\ohm}$, $L=L_s=\SI{100}{\mu H}$, $C_0=\SI{41.34}{pF}$, $C_1=-\SI{1.649}{pF/V^2}$, and $L_b=\SI{1000}{\mu H}$. There are $N = 30$ unit cells, and OBC is implemented by grounding both ends of the transmission line.  In the linear ($V_n \approx 0$) regime, each RLC resonator has natural frequency $\omega_0/2\pi = \SI{3.67}{MHz}$.  To apply the fundamental drive, we connect the PCB to three phase-synced function generators, which apply identical sinusoidal voltages of relative phase $\{0,\,2\pi/3,\,4\pi/3\}$ to sites $n \equiv \{0,\,1,\,2\}\pmod{3}$, respectively, consistent with the choice of wavenumber $k_p = 2\pi/3$ in the model. To activate the lossy BdG modes, we apply a weaker auxiliary drive at a tunable frequency different from the fundamental drive, with uniform amplitude and phase on all sites (adding phase differences does not substantially affect the results \cite{SM}). We then measure the voltage at each site and find its Fourier spectrum.

Fig.~\ref{fig:3_exp}(b) shows a typical spectrum obtained by the above procedure.  Here, we choose $V_{pp}=\SI{500}{mV}$ and $\omega_p/2\pi = \SI{3.81}{MHz}$ for the fundamental drive, so that the system is expected to lie well within the regime of broken pseudo-Hermiticity. For the auxiliary drive, we choose $V_{pp}' = \SI{50}{mV}$ and $\omega'/2\pi = \SI{3.88}{MHz}$ (i.e., normalized sideband frequency $\Omega = (\omega'-\omega_p)/\omega_p = 0.018$).  We observe a prominent sideband at precisely the auxiliary drive frequency, as well as a mirror peak on the opposite (undriven) side of the fundamental frequency.  The latter is expected due to the mixing between $\omega_f$ and $-\omega_f^*$ in the BdG ansatz \eqref{eq: BdG transformation}.

We repeat this procedure, changing the site $n$ where the voltage is measured, with all other parameters fixed.  In the upper panel of Fig.~\ref{fig:3_exp}(c), we plot the spatial distribution of the sideband amplitude $V_{n}^s \equiv \sqrt{|V_{n}^\ell|^2 + |V_{n}^r|^2}$, where $V_{n}^{\ell(r)}$ denotes the left (right) sideband voltage amplitude on site $n$.  It is found to be strongly localized to the right of the lattice.

If we probe the lattice at a larger auxiliary drive frequency ($\Omega = 0.058$), the sideband peaks remain observable but their spatial profiles are evenly distributed, as shown in the lower panel of Fig.~\ref{fig:3_exp}(c).  This is related to our previous theoretical finding that only a portion of the modes in the Bogoliubov spectrum may undergo pseudo-Hermiticity breaking [Fig.~\ref{fig:2_theory}(c)].  A more detailed comparison is given in Fig.~\ref{fig:3_exp}(d): the upper plot shows how the sideband amplitude varies with $n$ and $\Omega$, while the lower panel shows the theoretically-calculated Bogoliubov eigenfrequencies $\Omega_f$.  The frequency range where we experimentally observe strong localization is $\Omega \lesssim 0.03$, coincident with the range of $\mathrm{Re}(\Omega_f)$ where the Bogoliubov eigenfrequencies are non-real.

Moreover, the NHSE signature is strongly influenced by the driving amplitude.  In Fig.~\ref{fig:3_exp}(e), we plot the measured voltage spectra for three values of $V_{pp}$ (upper panels), and the corresponding sideband spatial profiles (lower panels), at $\Omega = 0.018$. When $V_{pp}$ is small, the NHSE disappears, along with left sideband. In fact, below a critical value of $V_{pp}$, there is no NHSE for any $\Omega$.  Hence, we can use the fundamental drive to perform active switching of the NHSE in the BdG modes.

To illustrate this effect, the upper panel of Fig.~\ref{fig:3_exp}(f) shows the effects of $V_{pp}$, for fixed $\Omega = 0.018$, on (i) the inverse participation ratio $\textrm{IPR} = \sum_n |V_{n}^s|^4/\left(\sum_m |V_{m}^s|^2\right)^2$, a measure of localization \cite{Thouless1974}, and (ii) the amplitude ratio between the left and right sidebands, $\lVert\mathbf{V}^\ell\rVert/\lVert\mathbf{V}^r\rVert$.  As we increase $V_{pp}$ past $\SI{240}{mV}$---previously identified theoretically to be the lattice's pseudo-Hermiticity breaking threshold [Fig.~\ref{fig:2_theory}(a)]---the experimental results show abrupt increases in the IPR (due to the NHSE) and the sideband amplitude ratio (due to the NH particle/hole pinning of the pseudo-Hermiticity-broken BdG eigenstates).  These experimental findings are also consistent with the detailed features of the theoretical BdG modes' wavefunctions.  In the lower panel of Fig.~\ref{fig:3_exp}(f), we plot the mean IPR and particle/hole ratio for the BdG modes, which evidently experience similar increases at around the same $V_{pp}$ threshold.  Note that the curves are smoothed out since they are averaged over all BdG modes, and not all BdG modes undergo symmetry breaking simultaneously.

\textit{Conclusions}---We have experimentally realized a classical nonlinear transmission line that hosts NH Bogoliubov quasiparticles.  Although there have been numerous demonstrations of tailored quasiparticle modes in linear synthetic metamaterials \cite{carusotto2013rmp, ozawa2019rmp}, including circuit lattices \cite{albert2015topological, ningyuan2015time, lee2018topolectrical}, Bogoliubov quasiparticles \cite{slim2024optomechanical, busnaina2024quantum} are qualitatively different because they are inherently nonlinear.  We have demonstrated their ability to host novel phenomena in the form of an actively switchable non-Hermitian skin effect, caused by the interplay between nonlinearity and spontaneous non-Hermitian symmetry breaking \cite{bender2007making, Wang2023tutorial}.  In the future, it would be interesting to use such methods to realize Bogoliubov quasiparticles that belong to other non-Hermitian topological classes \cite{kawabata2019symmetry}, which could exhibit more complex, and potentially useful, behaviors.

This work was supported by the Singapore National Research Foundation (NRF) under NRF Investigatorship NRF-NRFI08-2022-0001 and Competitive Research Program (CRP) NRF-CRP23-2019-0005, NRF-CRP23-2019-0007, and NRF-CRP29-2022-0003.

\bibliography{main}

\end{document}